\documentclass[aps,nofootinbib,showpacs,twocolumn]{revtex4-1}
\usepackage[CJKbookmarks, pdftex, bookmarksnumbered, bookmarksopen, colorlinks, citecolor=blue, linkcolor=blue]{hyperref}
\usepackage[english]{babel}
\usepackage{amstext,amsmath,amssymb,amsfonts,bbm}
\usepackage[latin1]{inputenc}
\usepackage{graphicx}
\usepackage{epstopdf}
\usepackage{amsthm}
\usepackage{tocvsec2}
\usepackage{enumerate}
\usepackage{subfigure}
\usepackage{color}

\begin{document}

\title{Statistical mechanics of reparametrization-invariant systems. It takes three to tango.}

\author{Goffredo Chirco}
\author{Thibaut Josset}
\author{Carlo Rovelli}
\affiliation{Aix Marseille Université, Université de Toulon, CNRS, CPT, UMR 7332, 13288 Marseille, France}

\date{January 21, 2016}

\begin{abstract} 
\noindent It is notoriously difficult to apply statistical mechanics to generally covariant systems, because the notions of time, energy and equilibrium are seriously modified in this context. We discuss the conditions under which weaker versions of these notions can be defined, sufficient for statistical mechanics. We focus on reparametrization-invariant systems without additional gauges.  The key is to reconstruct statistical mechanics from the ergodic theorem.  We find that a suitable split of the system into two non interacting components is sufficient for generalizing statistical mechanics. While equilibrium acquires sense only when the system admits a suitable split into \emph{three} weakly interacting components ---roughly: a clock and two systems among which a generalization of energy is equi-partitioned. This allows the application of statistical mechanics and thermodynamics as an additivity condition of such generalized energy.\end{abstract}

\maketitle



\section{Introduction}

Thermodynamics is a general tool for studying macroscopic phenomena in terms of a small number of concepts, such as energy, temperature, entropy, and equilibrium. Maxwell, Boltzmann, and Gibbs have related thermodynamics to the mechanical laws, and their statistical mechanics has been extended to quantum systems using Bose-Einstein and Fermi-Dirac distributions, and von Neumann entropy, which has lead to understanding phenomena such as black body radiation, superfluidity of $^4$He, or the stability of white dwarfs. So far, however, thermodynamics and statistical mechanics have not found a convincing generalization in a fully general relativistic context. 

The statistical mechanics of gravitationally interacting objects is already complicated in non relativistic approximation since gravity is long-range and attractive. But the problem is far deeper in the full relativistic theory, when we take into account the fact that thermal energy can be ceded to the degrees of freedom of spacetime itself. The difficulty comes from the modifications of the notion of \emph{time} introduced by general relativity \cite{Rovelli1995Analysis-of-the}.  In this context, no quantity concentrates all the properties that characterize the non relativistic notion of time on which statistical mechanics and thermodynamics are built. Neither coordinate time, nor clock times, nor proper time along a worldline fully behave as the time of non relativistic physics.
As a consequence, in a general covariant context there is no clear and unique notion of energy, with all the properties of Newtonian energy, to write statistical states or for thermodynamics. Another way of seeing the problem is to observe that if spacetime is a dynamical entity, a statistical state must describe also its thermal fluctuations --- in some sense, statistical mechanics must take fluctuations of time into account. 
 
An approach to the problem based on the idea of starting from arbitrary statistical states and deriving notions of time --thermal time-- and energy, from the state itself, has been developed in references \cite{Rovelli1993Statistical-mec,Connes1994Von-Neumann-alg, Rovelli2011Thermal-time-an, Rovelli2013General-relativ,Chirco2013Coupling-and-th}. However, this line of investigation has been delayed by a conceptual obstacle: the difficulty of characterizing the notion of \emph{equilibrium} within this perspective. 

Here we take a different path.  Standard statistical mechanics can be developed starting from the ``ergodic theorem", namely the observation that under suitable (``ergodic") conditions, the \emph{time average} of an observable is well approximated by an \emph{average over phase space}.  An equilibrium state is then defined as the phase-space probability measure 
satisfying this equality. This paper explores the possibility of founding covariant statistical mechanics on (an appropriate generalization of) this approach. More precisely, we extend the concept of time-average to reparametrization-invariant systems by means of a ``clock'', --a generalization of the notion of time--
and search for conditions under which the corresponding statistical state can be computed.

We develop this idea in the context of a reparametrization-invariant system with a finite --possibly large-- number of degrees of freedom and no other gauges beside coordinate-time reparametrization invariance. This is sufficient to show that a truly Newtonian time, with all the properties of time in the non relativistic theory, is \emph{not} necessary to use statistical mechanical methods. Generalization to field theory and systems with additional gauges is considered elsewhere. 

We report two main results. First, when the system admits a suitable split into two subsystems, the split itself allows defining a generalized time average and a statistical state. Roughly speaking, one subsystem can be taken as a ``clock system'' to describe thermal properties of the other. Statistical mechanics relies on this split, not on a peculiar choice of time variable in the clock system.

The structure defined above, however, is not sufficient to make sense of equilibrium. Our second result is that a split of the system into \emph{three} subsystems is
enough for having a notion of equilibrium. Roughly speaking: a clock-system and two systems in equilibrium with respect to each other. If the split satisfies certain conditions, the full machinery of equilibrium statistical mechanics can be used, and there will be an equilibrium configuration which maximizes entropy. The key condition is additivity: the split between system and clock defines a conserved quantity which is the sum of one component per  subsystem.

The paper is organized as follows. Section \ref{section_mechanics} reviews a formulation of Hamiltonian mechanics sufficiently wide to include general covariant systems, following Dirac \cite{Dirac1950Generalized-ham, Dirac1964Lectures-on-qua}, and Souriau \cite{Souriau1969Structure-des-s}. We call this generalized Hamiltonian formalism ``relativistic", following \cite{Rovelli2002A-note-on-the-f}. Section \ref{section_statistical_mechanics} introduces the notions of time average, statistical state, and ergodicity in this relativistic framework and defines the microcanonical ensemble when the system splits into two components. Section \ref{section_thermodynamics} derives the notion of equilibrium between two systems with respect to a third one, which yields to the definition of entropy, and suitable generalizations of the notions of energy, temperature, and time. In the Appendix \ref{flrw} we illustrate some aspects of the formalism by applying it to a radiation-dominated Friedmann-Lemaître-Robertson-Walker (FLRW) cosmology.


\section{Relativistic mechanics}\label{section_mechanics}

Classical mechanics describes how observables evolve in Newtonian time. This representation of dynamical phenomena is not possible for general relativistic systems. In this case, mechanics describes correlations between partial observables (which include time, or ``clock" observables), i.e., it describes how partial observables change with respect to one another. 


\subsection{Dirac's generalized Hamiltonian mechanics}

\begin{figure}[h] 
\includegraphics[width=2.5 in]{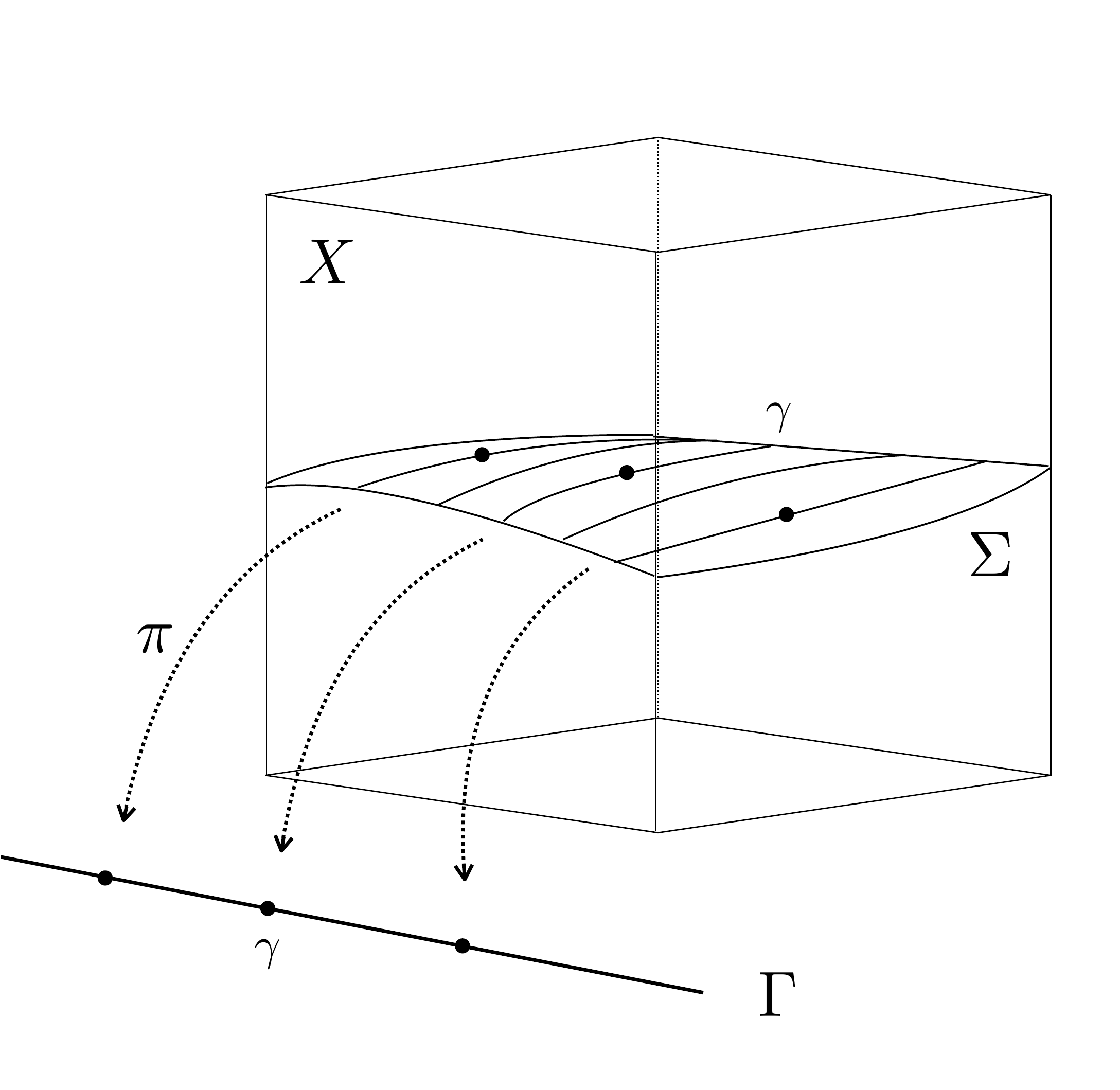}
\caption{Generalized hamiltonian mechanics}\label{generalized_hamiltonian_mechanics}
\label{fig_relativistic_mechanics}
\end{figure}

Systems like a free special-relativistic particle or the gravitational field in general relativity can be described by a manifestly covariant action, invariant under reparametrization of the evolution parameter. A Legendre transformation leads to the symplectic space $(X,\omega_X)$, called  the \emph{relativistic}, or \emph{extended}, \emph{phase space}, and a list of \emph{vanishing constraints} $\{C_i = 0 ~;~i=1..k\}$. The constraints define a \emph{presymplectic surface} $(\Sigma,\omega_\Sigma=\omega_X|_\Sigma)$, of codimension $k$. The dynamics of the system is fully encoded in the $k$-dimensional orbits in $\Sigma$, called \emph{states} (or \emph{motions}), which are the integral surfaces of the null directions of $\omega_\Sigma$. The set $\Gamma$ of these states is called \emph{physical phase space} and is naturally equipped with a symplectic structure defined by $\omega_\Sigma = \pi^* \omega_\Gamma$, where $\Sigma \overset{\pi}{\to} \Gamma$ (see figure \ref{fig_relativistic_mechanics}). For a detailed review of this presymplectic formalism, see \cite{Rovelli2002A-note-on-the-f,Rovelli2004Book}. 

In what follows, we restrict to systems with one-dimensional orbits ($k=1$), i.e., no gauge except reparametrization invariance. The orbits are the integral curves $\gamma$ of a vector field $Y$, on $\Sigma$, satisfying
\begin{equation}
	\omega_\Sigma (Y) = 0.  \label{eq_Y}
\end{equation}
Equivalently, $Y=Y_C|_\Sigma$, where $Y_C$ is the vector field, on $X$, satisfying the (generalized) Hamilton equations
\begin{equation}
	\omega_X ( Y_C ) = -dC \label{eq_hamilton}.
\end{equation}
Different constraints vanishing on the same surface define locally rescaled $Y$ and therefore different parametrizations of the \emph{same} orbits.  Equivalently, $Y$ is determined by \eqref{eq_Y} up to a local rescaling, which affects the parametrization of the orbits, and has no physical meaning. A state defines correlations between the partial observables (functions from $\Sigma$ to $\mathbb{R}$). The physics is fully coded by the presymplectic space $(\Sigma,\omega_\Sigma)$. 


\subsection{Non relativistic examples}\label{ex_non_relativistic}

Conventional Hamiltonian mechanics can be reformulated in this ``relativistic" language. A 
non relativistic system defined by a non relativistic phase space ($\Gamma_0,\omega_0$), a non relativistic Hamiltonian $H_0$, and Newtonian time $t$ is equivalent to a relativistic system defined by
\begin{equation*}
	\Sigma =  \mathbb{R} \times \Gamma_0 ~,~ \omega_\Sigma = -dH_0 \wedge dt + \omega_0.
\end{equation*}
 
\begin{figure}[h] 
\includegraphics[width=2.5 in]{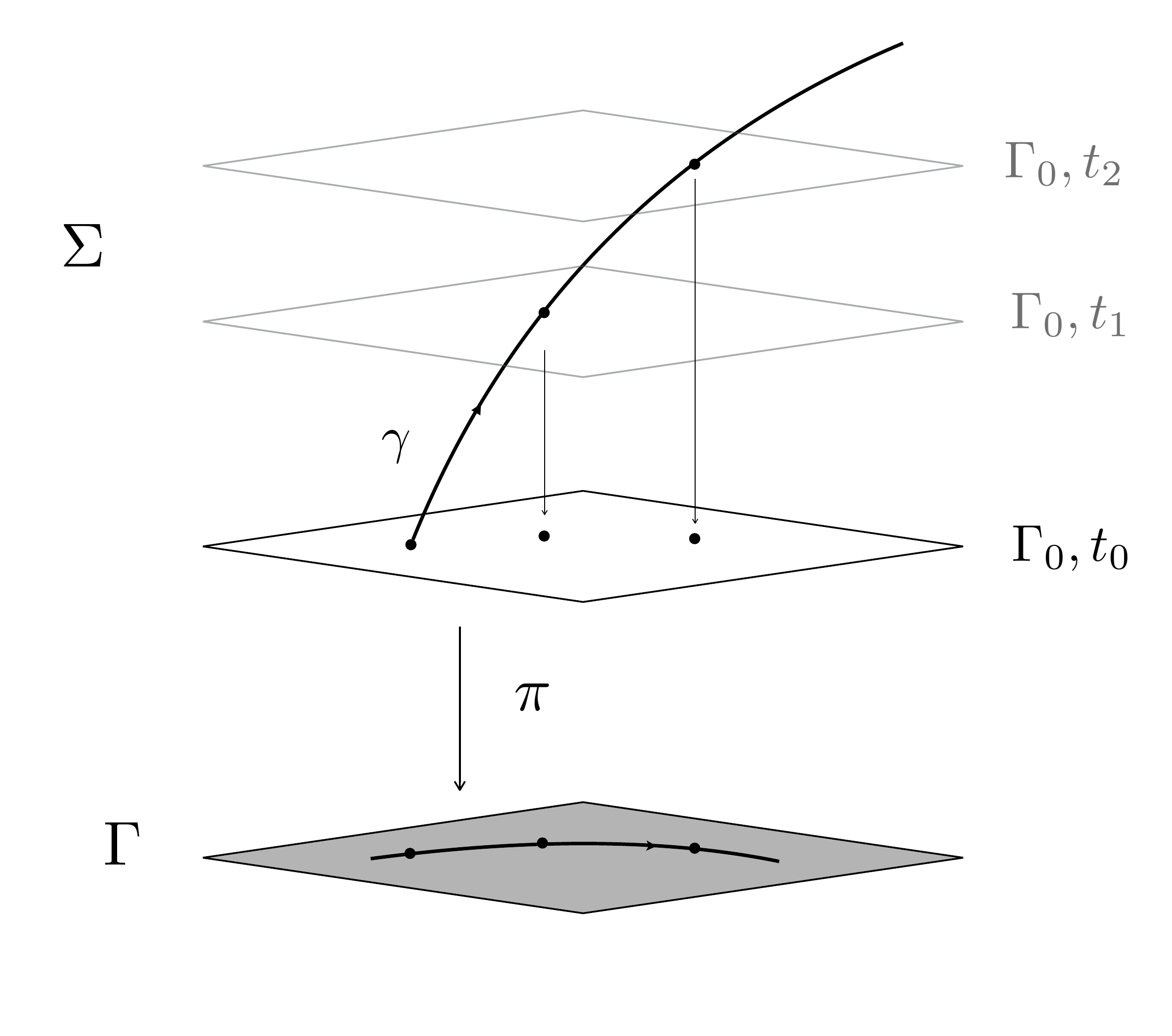}
\caption{Structure of the presymplectic surface of any non-relativistic system.}\label{non_relativistic_mechanics}
\label{fig_non_relativistic}
\end{figure}

By introducing a new variable $p_t$, as the conjugate momentum of $t$, one can also build an extended phase space 
\begin{equation*}
	X= T^*\mathbb{R} \times \Gamma_0 ~,~ \omega_X = dp_t \wedge dt + \omega_0,
\end{equation*}
and, with the constraint $C=p_t+H_0=0$, the equations of motion \eqref{eq_hamilton} reduce to usual Hamilton equations. The physical phase space $\Gamma$ is isomorphic to $\Gamma_0$, so
\begin{equation}\label{deparametrization}
	\Sigma \simeq \mathbb{R} \times \Gamma.
\end{equation}
This is called \emph{deparametrization} and allows dynamics to be interpreted as motion in phase space (Schrödinger's picture; see figure \ref{fig_non_relativistic}). 

There are relativistic dynamical systems that can be ``deparametrized", namely formulated also in the conventional non relativistic formalism. A free particle in Minkowski space $M$ is a typical example: $X=T^*M$ and $C=p^2+m^2=0$. Deparametrization requires breaking Lorentz symmetry, picking one Lorentz frame, where the constraint becomes $p_t+\sqrt{\vec p{\,}^2+m^2}=0$, and a Lorentz time $t=x^0$ as a special independent evolution parameter, even if physically there is nothing special about it. 

But there are also dynamical systems that do not admit a deparametrized version, namely that cannot be expressed in the conventional Hamiltonian language. An example is given below. 

\subsection{A relativistic example}\label{ex_relativistic}

Consider the system defined by 
\begin{eqnarray*} 
		X &=& \{a,p_a,\vec x,\vec p\} ~,~ \omega_X = dp_a \wedge da + d\vec p \wedge d\vec x \\
		C&=&\frac{1}{2}(p_a^2+a^2)+\frac{\vec p^2}{2m} + V(\vec x)=0,
\end{eqnarray*}
with a potential $V(\vec x)$ constraining $\vec x$ inside a finite volume. This systems looks similar to the classical description of a particle in a box evolving in Newtonian time. However, instead of the Newtonian time term $p_t$, there is the pendulum Hamiltonian $\frac{1}{2}(p_a^2+a^2)$, whose conjugate variable, the phase $\phi$ of the pendulum, is periodic: $\tan\phi=a/p_a$. The system describes the relative evolution of a particle in a box and a pendulum, but it cannot be deparametrized, because there is no globally monotonic time variable along the orbits: the orbits are closed.  This system admits no conventional Hamiltonian formulation. 

The question we address in this paper is whether statistical mechanics and thermodynamics methods can be extended to systems like this, that do not admit a Newtonian time parameter, or, more generally, that do not have a preferred time-evolution parameter.  The prime example of such systems is general relativity.


\section{Generalized Statistical Mechanics} \label{section_statistical_mechanics}

\subsection{Clock and time average}

In classical mechanics, we assume measurements to be instantaneous, to have arbitrary precision, and to not disturb the system. For a non relativistic system, an event is like a photo of all variables and their momenta $q^i(t),p_i(t)$, taken at a time $t$. The time average of a quantity $f$ is defined by
\begin{equation}\label{def_time_average}
	\bar f \underset{def}{=} \lim_{T\to +\infty} \frac{1}{2T} \int_{-T}^{+T} f(q^i(t),p_i(t))\,dt.
\end{equation}

If we treat the time variable $t$ as a partial observable on the same ground as the other variables, an event is labeled by $(q^i,p_i,t)$. In relativistic Hamiltonian formalism it is a point 
on the presymplectic surface $\Sigma$. Repeated measurements correspond to a list of events, all sitting on the same orbit $\gamma$.

We now want to generalize \eqref{def_time_average} to this context, i.e., we want to define the \emph{average} of an observable $f$ over a list of events.  But there is no meaning in taking averages along an orbit without a measure on it.

A first crucial step is the observation that a measure on the orbit does \emph{not} require a time variable $t$ to exist. A one-form generalizing $dt$ is required, and it suffices. Let us therefore call ``clock" a one-form $\theta$ on $\Sigma$ that defines a volume form along each orbit.  For any quantity $f:\Sigma \to \mathbb{R}$ and any state $\gamma$, we can then define the \emph{$\theta$-average of $f$ along $\gamma$} by
\begin{equation} \label{def_theta_average}
	\bar f(\theta,\gamma) \underset{def}{=} \frac{\int_\gamma f \theta}{\int_\gamma \theta}.
\end{equation}

In the simple example in the section \ref{ex_relativistic}, the phase of the pendulum determines a clock $\theta=\frac{p_a da -a dp_a}{a^2+p_a^2}$ which is well defined on the orbit even if $\theta=d\phi$ makes sense only locally, and there is no global time parameter along the orbit.  The operational procedure for measuring the average \eqref{def_theta_average} is well defined.

\subsection{Statistical state}

Given a clock $\theta$ and a state $\gamma$, $\theta$-averaging along $\gamma$ is a linear functional
\begin{equation*}
	f \mapsto \bar f(\theta,\gamma) \in \mathbb{R}.
\end{equation*}
We call \emph{$\theta$-statistical state} the (probability) measure $\mu_{\theta,\gamma}$ over $\Sigma$\footnote{If one restricts to functions $f:A \to \mathbb{R}$ only (where, for example, $A \subset \Sigma$ or  there is a projection $\Sigma \to A$) then the statistical state $\mu_{\theta,\gamma}$ is a measure on $A$.} naturally defined by
\begin{equation}\label{def_statistical_state}
	\int_\Sigma f d\mu_{\theta,\gamma} \underset{def}{=} \bar f (\theta,\gamma).
\end{equation}
So far the statistical state $\mu_{\theta,\gamma}$ is nothing more than another name for $\theta$-average along $\gamma$. But in the next paragraph, we show how this statistical state can be computed without solving the dynamics, if the system is, in some appropriate sense, ergodic.


\subsection{Ergodicity}

Ergodicity can be understood in various manners\footnote{As in the classical works of Maxwell \cite{Maxwell1860Illustrations-o}, Boltzmann \cite{Boltzmann1896Lectures-on-Gas}, von Neumann \cite{Neuman1931Proof-of-the-qu,Neuman1932Physical-applic}, Birkhoff \cite{Birkhoff1931Proof-of-the-er} and Khinchin \cite{Khinchin1949Mathematical-fo}.}.  A general framework for expressing it is the following: a measured dynamical system $(V,U_t,\nu)$\footnote{$(V,\nu)$ is a probability space and $U_t : V \to V$ is a measure-preserving time evolution operator.} is ergodic if for almost any $P_0 \in V$,
\begin{equation}\label{ergodic_thm}
	\lim_{T \to +\infty} \frac{1}{2T} \int_{-T}^{+T} f(U_t P_0)\, dt = \int_{V} f\, d\nu.
\end{equation}

It is common to say that statistical mechanics is valid when the ``ergodic hypothesis'' is satisfied. Concretely, ergodicity is notoriously hard to prove mathematically and notoriously very common in nature. Here we are not interested in \emph{proving} ergodicity in a general covariant context, but in understanding how it can be \emph{formulated} in such a context.

Intuitively, \eqref{ergodic_thm} means that (almost) any history explores $V$ entirely and therefore the time average does not depend on the initial conditions $P_0$. Then there is a natural way of extending the notion of ergodicity to the relativistic hamiltonian context: First, we say that a partial observable $f$ on $\Sigma$ is \emph{$\theta$-ergodic} if its $\theta$-average does not depend on the state $\gamma$ of the system, i.e.,
\begin{equation}\label{def_ergodic_function}
	\bar f(\theta,\gamma) = \bar f (\theta) \qquad \forall \gamma \in \Gamma.
\end{equation}
Then, a system is called $\theta$-ergodic if \emph{every} partial observable is $\theta$-ergodic. Equivalently, if the $\theta$-statistical state is independent on the (pure) state $\gamma$ of the system, i.e.,
\begin{equation}\label{def_ergodic_system}
	\mu_{\theta,\gamma} = \mu_\theta \qquad \forall \gamma \in \Gamma.
\end{equation}

In Appendix \ref{proof_ergodic_statistical_state}, we prove that if the relativistic system $(X,\omega_X,C)$ is $\theta$-ergodic, then the statistical state $\mu_\theta$ on $\Sigma$ is given by the explicit formula
\begin{equation}\label{ergodic_statistical_state}
	d\mu_\theta = \frac{\theta(Y_C)\, \delta(C)\, d\mu_X}{\int \theta(Y_C)\, \delta(C)\, d\mu_X},
\end{equation}
where $Y_C$ is defined in \eqref{eq_hamilton}. 

An essential point here is that in spite of the appearance of $C$ in this expression, this statistical state does not depend on the choice for the constraint $C$: it is entirely 
determined by the presymplectic structure. In this sense, it is invariant under the reparametrization gauge.  On the other hand, it depends on the choice of the clock. In the next section we see how this choice can be physically motivated.


\subsection{Microcanonical ensemble}

Thus far, the clock has been taken to be arbitrarily chosen. Different clocks lead to different statistical mechanics. We now focus on a situation when this arbitrariness is reduced and the choice of the clock has a clear physical ground.   

The key requirement we demand to a realistic physical clock is to interact as little as possible with the system we measure.  If we are interested in the average temperature on Earth, we better not have a clock that shuts itself off during summer.  In a general covariant context, this requirement is captured by demanding that the global system splits into two non interacting parts---one of which playing the role of ``clock system"---in the following sense.

A mechanical system $\mathcal{S}$, given by $(\Sigma,\omega_\Sigma)$, \emph{splits into two non-interacting subsystems} $\mathcal{S}^a$ and $\mathcal{S}^b$, if it can be written as the presymplectic surface in an extended phase space $(X=X^a\times X^b, \omega_X = \omega_{X^a} + \omega_{X^b})$, defined by a constraint of the form
\begin{equation*}
C= C^a+C^b=0,
\end{equation*}
where $C^\alpha : X^\alpha \to \mathbb{R}$ ($\alpha=a,b$). 

Such a split determines a foliation of the presymplectic surface 
\begin{equation}\label{foliation_presymplectic}
	\Sigma = \bigsqcup_{I^a+I^b=0} \Sigma^a_{I^a} \times \Sigma^b_{I^b},
\end{equation}
where $X^\alpha= \bigsqcup_{I^\alpha} \Sigma^\alpha_{I^\alpha}$ and $I^\alpha$ is the value of $C^\alpha$ on the leave.  Each $\left( \Sigma^\alpha_{I^\alpha}, \omega_{X^\alpha}|_{\Sigma^\alpha_{I^\alpha}} \right)$ is the presymplectic space that would describe the subsystem $\mathcal{S}^\alpha$ if it was isolated, and we can easily check that
\begin{equation*}
	\dim \Gamma =  \dim \Gamma^a_{I^a} + \dim \Gamma^b_{I^b} + 2.
\end{equation*}
The degrees of freedom of $\mathcal{S}$ are those of $\mathcal{S}^a$, plus those of $\mathcal{S}^b$, plus one degree of freedom at the boundary of the two subsystems.

\paragraph*{Example.}{
According to \ref{ex_non_relativistic}, any non-relativistic system with $N$ degrees of freedom splits into a time part 
\begin{eqnarray*}
	X^{a} &=& \{t,p_t\} ~,~\omega^{a} = dp_t \wedge dt ~,\\
	C^{a} &\underset{def}{=}& p_t=I^{a}
\end{eqnarray*}
and the non relativistic phase space  
\begin{eqnarray*}
	X^b &=& \{q^i,p_i\} ~,~ \omega^b = dp_i \wedge dq^i ~,\\
	C^b &\underset{def}{=}& H_0(q^i,p_i)=I^{b},
\end{eqnarray*}
where $I=I^b=-I^a\in \mathbb{R}$ is the standard energy. 

Notice that the time part alone has zero degrees of freedom, and it admits a single orbit: the absolute flow of Newtonian time, which appears as a gauge. The other part has $N-1$ degrees of freedom: the dynamics at fixed energy, again with evolution as a gauge.  By coupling the two systems, a novel physical degree of freedom arises, describing the \emph{relative} evolution of the two gauge parameters \cite{Rovelli2014Why-Gauge}, i.e., the coupling introduces the evolution of the system with respect to the Newtonian time $t$.}

\paragraph*{Example.}{
The relativistic system \ref{ex_relativistic} splits into a pendulum component
\begin{eqnarray*}
	X^{a} &=& \{a,p_a\} ~,~ \omega_{X^{a}} = dp_a \wedge da,\\
	C^{a}&\underset{def}{=}& \frac{1}{2}(p_a^2+a^2) =I^a
\end{eqnarray*}
and a particle component
\begin{eqnarray*}
	X^{b} &=& \{\vec x, \vec p\} ~,~ \omega_{X^{part}} = d\vec p \wedge d\vec x ,\\
	C^{b}&\underset{def}{=}& \frac{\vec p^2}{2m} + V(\vec x) =I^b.
\end{eqnarray*}
Once again, the pendulum part $(X^{a}, C^{a})$ alone defines a system with no degrees of freedom, as it has a single orbit. The coupled system does not describe the partial observables of the particle as functions of absolute Newtonian time, but instead their correlations with pendulum variables.
}

Now we return to statistical mechanics.  The Newtonian time subsystem in the first example and the pendulum in the second are special cases of what we call a ``clock subsystem". In general, however, a clock subsystem does not need to be one-dimensional.   

Say we are interested in measuring partial observables of the subsystem $\mathcal{S}^b$ only (i.e., $f\!:X^b \to \mathbb{R}$), using a clock in the subsystem $\mathcal{S}^a$ (i.e., \,$\theta$ is a $1$-form over $X^a$). Then, assuming $\theta$-ergodicity, we get a family of statistical states, labeled by the generalized energy $I=I^b=-I^a$, defined on $\Sigma^b_{I}$. In Appendix \ref{proof_microcanonical}, we prove that this state is given explicitly by 
\begin{equation}\label{microcanonical_state}
	d\mu_{\Sigma^b_I} = \frac{\delta(C^b-I)\, d\mu_{X^b}}{\int \delta(C^b-I)\, d\mu_{X^b}}.
\end{equation}
These states form the \emph{microcanonical ensemble} for the subsystem $\mathcal{S}^b$. 

Remarkably, these states depends only on the splitting of $\Sigma$. They do not depend on the specific choice of clock one-form on $\mathcal{S}^a$, neither on the form of the constraint.

This is our first main result: microcanonical states can be defined in a context much wider than conventional Hamiltonian mechanics.  They can be defined for any relativistic Hamiltonian system admitting a split into two non interacting components, and they depend only on such a split. Assuming ergodicity they allow to express averages along the orbits (which require solving the dynamics) in terms of phase-space averages (which do not require solving the dynamics).  


\section{Generalized Thermodynamics}\label{section_thermodynamics}

The three laws of thermodynamics are \emph{all} in question, in a general covariant context. Energy conservation requires energy to be defined, but in a generic general covariant system there is no non trivial conserved energy: a covariant generator of evolution along the orbit is the constraint, but it vanishes identically. The second law is tied to a notion of equilibrium and time variable, which, as we see below, requires more structure to be defined. Finally, the ``zero-th" law, namely the equality of temperature at equilibrium appears to be violated already for systems in a fixed relativistic background, as with the Tolman-Ehrenfest effect \cite{Tolman}.  Under which conditions do we recover thermodynamics in a reparametrization-invariant system? 

\subsection{First law}

We have seen that a quantity $I$, conserved along the evolution, exists if the covariant system splits into two non interacting subsystems. As we saw in the example above, $I$ can be identified with the conventional energy of the non relativistic formulation, if this exists.  Therefore $I$ is a generalization of the notion of energy.  Given a split, the first law generalizes immediately to $I$ conservation. 


\subsection{Zero'th law}

Consider two pieces of metal at different temperatures. Put them into contact and isolate the full system. Thermodynamics predicts that energy will flow between the two, until equilibrium is reached. Entropy measures how well shared is the total energy between the two pieces. How do we recover this prediction in the relativistic Hamiltonian context?  

Notice that time does not play a significant role. What matters is the existence of a quantity (energy) whose total amount is fixed and which can be shared between subsystems. In section \ref{section_statistical_mechanics} we showed that a split of a system into two non interacting components was sufficient to define a statistical state for one component, but a split into three (here the two pieces of metal and Newtonian time) is required in order to talk about equilibrium between subsystems.

For a general covariant system, assume that $\mathcal{S}$ splits into two subsystems $\mathcal{S}^{a}$ and $\mathcal{S}^{bc}$. Choose a clock in $\mathcal{S}^a$ 
and assume also that $\mathcal{S}^{bc}$ is in turn comprised of two interacting subsystems $\mathcal{S}^b$ and $\mathcal{S}^c$, i.e.,
\begin{eqnarray*}
	X &=& X^a \times X^b \times X^c, \\
	\omega_X &=& \omega_{X^a} + \omega_{X^b} +\omega_{X^c}, \\
	C &=& C^a + C^b+ C^c + V^{bc} =0, \\
	\Sigma &=& \bigsqcup_{I^{a}+I^{bc}=0} \Sigma^{a}_{I^{a}} \times \Sigma^{bc}_{I^{bc}}, 
\end{eqnarray*}
where $V^{bc}$ depends on both $X^b$ and $X^c$ but not on the clock subsystem $X^a$, and represents an interaction terms that allows (a suitable generalization of) energy to flow between the two subsystems $\mathcal{S}^b$ and $\mathcal{S}^c$.

Assuming ergodicity for the isolated system $\Sigma^{b,c}_{I^{bc}}$, the microcanonical state \eqref{microcanonical_state} is given by
\begin{equation*}
	d\mu_{\Sigma^{bc}_{I^{bc}}} = \frac{\delta(C^b+C^c+V^{bc} - I^{bc})\, d\mu_{X^b \times X^c}}{\int \delta(C^b+C^c+V^{bc} - I^{bc})\, d\mu_{X^b \times X^c}}.
\end{equation*}

We can now apply the usual machinery of statistical mechanics to define equilibrium between the two systems $\mathcal{S}^b$ and $\mathcal{S}^c$, i.e., if $V^{bc}$ is a small perturbation (Boltzmann's thermalizing ``grain of sand") then
\begin{equation*}
	\delta(C^b+C^c+V^{bc} - I^{bc})d\mu_{X^b \times X^c} \approx \delta(C^b+C^c - I^{bc}) d\mu_{X^b \times X^c}
\end{equation*}
and the microcanonical state factorizes
\begin{equation*}
	d\mu_{\Sigma^{bc}_{I^{bc}}} \propto [ \delta(C^b-I^b) d\mu_{X^b} ] [  \delta(C^c-(I^{bc}-I^b)) d\mu_{X^c} ]  dI^b .
\end{equation*}
$I^b$ and $I^c$ are not conserved, but they sum up to the total conserved $I$-{\em energy} quantity $I=I^{bc}=I^b+I^c$. 
The probability distribution of having $I$-energy partitioned as $(I^b,I^c)$, is given by
\begin{equation*}
	\Omega(I^b,I^c) \propto \Omega^b(I^b)\  \Omega^c(I^c),
\end{equation*}
where $\Omega^\alpha(I^\alpha) =  \int \delta(C^\alpha-I^\alpha) d\mu_{X^\alpha}$ ($\alpha=b,c$) is the ``phase space volume" of the $I$-energy surface. 

We can then define, for each subsystem, its entropy
\begin{equation}\label{I_entropy}
	S(I) \underset{def}{=} k_B \log \Omega(I)
\end{equation}
and its \emph{I-temperature}
\begin{equation}\label{I_temperature}
	\frac{1}{T_I} \underset{def}{=} \frac{dS(I)}{dI}.
\end{equation}

The \emph{equilibrium} value of $(I^b,I^c)$ is obtained by maximizing the entropy of the isolated system $\mathcal{S}^{bc}$
\begin{equation*}
S(I^b,I^c) = S^b(I^b)+S^c(I^c),
\end{equation*}
or, equivalently, by equal $I$-temperatures
\begin{equation*}
T_I^a|_{eq}=T_I^b|_{eq}.
\end{equation*}
The $0^{th}$ law of thermodynamics is recovered in term of $I$-temperatures. 

A conventional hamiltonian system written in the relativistic formalism (section \ref{ex_non_relativistic})  splits into a time part and the system itself. For such a system, \eqref{microcanonical_state} reduces to the usual microcanonical state of non relativistic system.  If the system splits into two weakly interacting subsystems, i.e., $H_0\sim H_0^b+H_0^c$, then \eqref{I_entropy} and \eqref{I_temperature} are the usual entropy and temperature. 

But in general the $I$-temperature is not necessary what would be measured by a standard thermometer.  A good example of this discrepancy is the Tolman-Ehrenfest effect: the temperature $T$ measured by a standard thermometer in an equilibrium configuration of matter on a fixed static gravitational field is \emph{not} constant in space.  The reason is that $T$  is the derivative of the entropy with respect to the local matter energy, which is the generator of proper-time translations, and is not conserved, because of the interaction with gravity.  A conserved quantity $I$ is given in this case by the generator of the global time translations in the static gravitational field. The ratio between $T$ and $T_I$ is different in different space points, because the ratio between the global time and proper time is affected by the local gravitational field. The general formalism developed here brings back the Tolman-Ehrenfest's apparent violation of the zero-th principle into standard (but generalized) thermodynamics.  

\subsection{Second law}

Here we are only concerned with the problem of extending the second law to the relativistic case, and not to address the numerous subtleties and questions raised by the law.

The physics of the second law is captured by the following fact: for most states that at time $t_0$ have an entropy $S(t_0)$ lower than the maximal entropy, dynamics takes the system to a state with higher entropy at time $t_0+dt$. (For most states, $S(t_0-dt)$ is larger than $S(t_0)$ as well.) This behavior is obtained as soon as the system is ergodic and the probability distributions of macroscopic observables are very peaked. 

Remarkably, the width of probability distributions is simply related to the notion of subsystem: if the system is macroscopic (i.e., can be split into a very large number of weakly interacting subsystems) then we can expect the second law to hold. This is a physical consequence of the law of large numbers and has nothing to do with the relativistic/non relativistic aspects of the system. 

Thus, the second law generalizes immediately to the relativist context: entropy increases along the evolution for most low-entropy configurations of a reparametrization-invariant system formed by a large number of weakly interacting components.

\subsection{Gauge invariance and additivity}

We have seen that the existence of a split of a reparametrization-invariant system into two non interacting subsystems implies the existence of a conserved quantity $I$. It is important to note, however, that this quantity is defined by the split only up to reparametrization $I\to \tilde I=f(I)$. 

Recall, indeed, that $I$ is defined as the value of $C^b$ when the total constraint has the form $C^b=-C^a$. But the same dynamics can be obtained from the constraint $f(C^b)=f(-C^a)$ for a rather generic (bijective) function $f$. Therefore $\tilde I=f(I)$, which is of course a conserved quantity as well, can be equally well taken as a generalization of the notion of energy.   

Accordingly, we can use $C^b$ to define an evolution parameter $\tau$ along the orbits by $dA/d\tau=\{A,C^b\}$, but this choice is gauge dependent. The split between clock and the rest is sufficient to define the microcanonical states, but is not sufficient to determine a gauge-invariant preferred generalized time variable.  In other words, the statistical mechanics we defined in section \ref{section_statistical_mechanics} does \emph{not} require a preferred notion of time. 

The situation changes when we consider a split into \emph{three} subsystems and need all systems are weakly interacting.   The reason is that if we now replace the generalized energy $I=C^{bc}$ with $\tilde I=f(I)$, its additivity, $I=I^b+I^c$, is lost, unless $f$ is affine.  Therefore the $I$-energy is now defined only up to affine transformations: changes of units and energy zero point.  

The crucial property for the definition of equilibrium, therefore, is the additivity of the conserved quantity that exists if the system admits a three-way split. 

If we now use the generalized energy $I=C^{bc}$ to define an evolution parameter $\tau$ along the orbits by $dA/d\tau=\{A,C^{bc}\}$, the resulting evolution parameter $\tau$ is no longer gauge dependent: it is the unique parameter (up to rescaling) whose conjugate energy is additive.   (A moment of reflection shows that $\tau$, defined in this manner, is precisely the thermal time of the equilibrium state \cite{Rovelli1993Statistical-mec,Connes1994Von-Neumann-alg, Rovelli2011Thermal-time-an, Rovelli2013General-relativ,Chirco2013Coupling-and-th}.)

\section{Conclusion}

We studied the conditions under which concepts and tools of statistical mechanics and thermodynamics extend to reparametrization-invariant systems. 

To this end, we studied under which conditions suitable generalizations of the notions of time, energy, and equilibrium can be found, sufficient for statistical mechanics and thermodynamics to become meaningful.  

The key idea was to study what is needed to extend the notion of ergodicity (i.e., suitable independence of averages from initial conditions) to a reparametrization-invariant system.  

We introduced the following main ideas:
\begin{description}
\item[\rm\em Clock] Non relativistic time sets the $dt$ measure needed for time averages. In a general covariant context, the clock describes a physical devise measuring averages along the evolution.  
\item[\rm\em Split into two subsystems] A split of the system into two non interacting parts is sufficient for determining a class of equivalent clocks: one-forms depending only on the clock subsystem define the same statistical state of the rest of the system. The split into two non interacting parts also implies the existence of a conserved quantity, denoted $I$-energy. 
\item[\rm\em Split into three weakly interacting parts] If the system splits into \emph{three} (or more) weakly interacting parts, equilibrium can be defined by maximizing the entropy under $I$-energy exchange. $I$-energy additivity picks a preferred evolution parameter that can be identified with the thermal time of the equilibrium state.
\item[\rm\em $I$-temperature] $I$-temperature (in general distinct from conventional temperature) is the (inverse) derivative of entropy by $I$-energy and is constant at equilibrium. 
\end{description}

We showed that a split of the system into two non interacting parts is sufficient to define the microcanonical ensemble and therefore to use the tools of statistical mechanics.

A split of the system into three (or more) weakly interacting parts is needed for useful notions of equilibrium and temperature, and is sufficient for deriving a generalization of all three laws of thermodynamics. 

In Appendix \ref{flrw} we illustrate the formalism in the context of the reparametrization-invariant system describing a homogeneous FRLW cosmology filled with non homogeneous electromagnetic radiation.  We show that the equilibrium state is characterized by an $I$-temperature $T_I$ constant in time. The conventional temperature $T$ is related to  $T_I$ by a time-dependent scaling factor, giving the right time dependence of the radiation-filled FRLW cosmology.

The formalism introduced in the paper tackles the \emph{conceptual} issue of understanding thermodynamical properties in the absence of Newtonian time and allows a \emph{covariant} treatment of simple relativistic systems, such as the cosmological example in Appendix \ref{flrw}. Is it relevant for gravitational physics? Investigating statistical mechanics of more involved, realistic, gravitational systems necessarily require more work: one should first consider the present approach for systems with more than one constraint (more gauge symmetries) and then extend it to a fully relativistic field-theoretic framework.

Nevertheless, the scheme was derived to be well suited for a generalization to a multisymplectic or covariant description of reparametrization-invariant field theories: within this framework, mechanics of relativistic fields, on a 3+1-dimensional manifold, can be formulated in terms of a finite dimensional surface $\Sigma$, equipped with a 5-form, leading to 4-dimensional orbits (see e.g. \cite{,Rovelli2004Book}). The 1-form used to define a clock will be replaced by a 4-form, providing a (gauge-invariant) measure on the orbits; the strategy would then consist in rethinking the notion of ergodicity and the derivation of statistical state in this context.

If this is possible, the notions introduced here could perhaps be used to study the statistical mechanics and the thermodynamics of the gravitational field. In this context, the relevant split could be between regions of spacetime, or between spacetime and some matter, or else.  As always in physics, \emph{approximate} lack of interaction between clock systems and the other (partial) observables should be sufficient.

\subsection*{Acknowledgements}

We are grateful to Alejandro Perez and Simone Speziale for the interesting discussions and insights.

\appendix

\section{Statistical state} \label{proof_ergodic_statistical_state}

In this appendix we derive the form of the statistical state \eqref{ergodic_statistical_state}. To make the proof rigorous, some additional mathematical hypotheses are required: the $1$-form $\theta$ to be with compact support, the functions $f$ to be continuous with compact supports and ``$\delta(C) d\mu$'' to be understood as the measure defined by the disintegration theorem.
	
Let $\tau$ be the parameter generated by $C$ along the orbits, so that $Y_C=d/d\tau$. Let 
 $\{q^\mu, p_\mu\}$ be coordinates on $X$ such that $\omega_X = dp_\mu \wedge dq^\mu$. Then $Y_C = \frac{\partial C}{\partial p_\mu} \frac{\partial}{\partial q^\mu}-\frac{\partial C}{\partial q^\mu} \frac{\partial}{\partial p_\mu}$. Writing $\theta = \alpha_\mu dq^\mu + \beta^\mu dp_\mu$, one can check that, for any orbit $\gamma$, 
\begin{equation*}
	\gamma^* \theta = \left( \alpha_\mu \frac{\partial C}{\partial p_\mu} - \beta^\mu \frac{\partial C}{\partial q^\mu} \right) d\tau = \theta (Y_C)|_\gamma \, d\tau.
\end{equation*}
Thus, along each orbit $\theta (Y_C)$ determines the ratio between the clock measure and the measure determined by the (gauge dependent) evolution generated by $C$.  

$\delta(C) d\mu_X$ defines a measure on $\Sigma$, invariant under the (unphysical) time evolution induced by $C$, i.e., $\forall g$,
\begin{eqnarray*}
	& & \int g[P(P_0,\tau)] \delta \left[ C(P_0) \right] d\mu_X[P_0]  \\
		&& \ \ \ = \int g[P_0] \delta\left[C(P_0)\right] d\mu_X [P_0].
\end{eqnarray*}

We denote $\gamma_{P_0}$ the (unique) orbit going through a point $P_0\in \Sigma$. $\theta$-ergodicity for a function $f$ \eqref{def_ergodic_function} can be written
\begin{equation*}
	\int_{\gamma_{P_0}} f\theta = \bar f(\theta) \int_{\gamma_{P_0}} \theta ~,~\forall P_0.
\end{equation*}
So, by definition,
\begin{equation*}
	\int_\mathbb{R} \left( \gamma_{P_0}\right)^*[f\theta] = \bar f(\theta) \int_\mathbb{R} \left(\gamma_{P_0}\right)^* [\theta] ~,~\forall P_0,
\end{equation*}
thus $\forall P_0 \in \Sigma$, with obvious notation, 
\begin{equation*}
	\int f[P(P_0,\tau)] \, \theta (Y_C)|_{P(P_0,\tau)} \, d\tau = \bar f(\theta) \int \theta (Y_C)|_{P(P_0,\tau)} \, d\tau.
\end{equation*}
By integrating both sides with the preserved measure $\delta[C(P_0)] d\mu_X[P_0]$, we get
\begin{equation*}
	\Delta \tau \int f \, \theta (Y_C) \,  \delta(C) d\mu_X = \bar f(\theta) \Delta \tau \int \theta (Y_C) \,  \delta(C) d\mu_X,
\end{equation*}
This gives 
\begin{equation*}
	\bar f(\theta) = \frac{ \int f \, \theta (Y_C) \, \delta(C)\, d\mu_X}{\int \theta (Y_C) \,  \delta(C)\, d\mu_X},
\end{equation*}
from which  \eqref{ergodic_statistical_state} follows immediately.  The $\theta (Y_C)$ factors compensates the scaling of the delta function under a change of $C$, so that the state is gauge invariant.

\section{Microcanonical ensemble}\label{proof_microcanonical}

In this section we derive the microcanonical ensemble \eqref{microcanonical_state}. Let's write 
\begin{equation*}
C=C^a+C^b=0 ~,~ \Sigma = \!\!\bigsqcup_{I^a+I^b=0} \Sigma^a_{I^a} \times \Sigma^b_{I^b}.
\end{equation*}

$I=I^b=-I^a$ is conserved along each orbit. According to Birkhoff's ergodic theorem, it is known that ergodicity can be satisfied, at most, on slice of constant $I$, so $\theta$-ergodicity for a function $f:X^b \to \mathbb{R}$ should be understood as
\begin{equation*}
\bar f(\theta,\gamma) = \bar f(\theta,I) ~,~\forall \gamma \subset \Sigma_{I}=\Sigma^a_{-I} \times \Sigma^b_{I}.
\end{equation*}
Then, $\forall P_0 \in \Sigma_I$,
\begin{equation*}
	\int f[P(P_0,\tau)] \, \theta (Y_C)|_{P(P_0,\tau)} \, d\tau = \bar f(\theta,I) \int \theta (Y_C)|_{P(P_0,\tau)} \, d\tau.
\end{equation*}
By integrating both sides with the preserved measure $\delta[C^b(P_0)-I] \delta[C(P_0)] d\mu_X[P_0]$, we get
\begin{equation*}
	\bar f(\theta,I) = \frac{\int f\,  \theta (Y_C)\, \delta(C^b-I)  \delta(C)\, d\mu_X}{\int  \theta (Y_C) \, \delta(C^b-I) \,  \delta(C)\, d\mu_X}.
\end{equation*}

The split leads to
\begin{eqnarray*}
	d\mu_X &=& d\mu_{X^a} d\mu_{X^b}, \\
	\delta(C) &=& \delta(C^a+C^b), \\
	Y_C &=& Y_{C^a} + Y_{C^b}.
\end{eqnarray*}
As the clock $\theta$ depends only on the partial observables of the subsystem $\mathcal{S}^a$, $\theta(Y_{C^b})=0$. In addition, the function $f$ depends only on the partial observables of the subsystem $\mathcal{S}^b$ so the integral
\begin{equation*}\,
	\int f\, \theta (Y_C) \, \delta(C^b-I)  \delta(C) \, d\mu_X
\end{equation*}
factorizes into
\begin{equation*}
	\left( \int  \theta(Y_{C^a})\, \delta(C^a+I) \, d\mu_{X^a} \right) \left( \int  f\, \delta(C^b-I) \, d\mu_{X^b} \right).
\end{equation*}
Finally, taking the normalization factor into account, 
\begin{equation*}
	\bar f(\theta,I) = \frac{\int f\delta(C^b-I) d\mu_{X^b}}{\int \delta(C^b-I) d\mu_{X^b}},
\end{equation*}
which implies immediately \eqref{microcanonical_state}.


\section{An application: electromagnetic field in an expanding universe}\label{flrw}

We illustrate some of the notions that we have introduced by studying a simple infinite-dimensional relativistic system: a FLRW universe filled with a non homogeneous electromagnetic field (cosmic radiation) \cite{Rovelli1993The-statistical}. The dynamics is not given by the full Einstein equations since only the scale factor $a(t)$ interacts with the electromagnetic field, the other degrees of freedom of the gravitational field are assumed to be frozen in the homogeneous configuration. The aim is to describe the statistics of the cosmic radiation. 

The system is described by the single constraint
\begin{equation*}
	C = - \frac{2\pi G}{3 V } \frac{p_a^2}{a} - \frac{3V\kappa}{8\pi G}a + H = 0
\end{equation*}
and the symplectic form
\begin{equation*}
	\omega = dp_a \wedge da + \int \left( dE^i(\vec x) \wedge dA_i(\vec x) \right) d^3\vec x,
\end{equation*}
where
\begin{eqnarray*}
	ds^2 &=& -dt^2 + a(t)^2 \tilde g_{ij} dx^i dx^j, \\
	V &=& \int \sqrt{\tilde g} d^3 \vec x, \\
	H &=& \int T_{00}[g,E,A] \sqrt{g} d^3 \vec x = \frac{1}{a} \int \tilde H(\vec x) d^3 \vec x,\\
	\tilde H (\vec x) &=& \tilde g^{-1/2} \tilde g_{ij} E^i E^j + \tilde g^{1/2} \tilde g^{ik} \tilde g^{jl} \partial_{[i} A_{j]} \partial_{[k} A_{l]}.
\end{eqnarray*}
Assuming flat space ($\kappa=0$, $\tilde g_{ij}=\delta_{ij}$) and performing Fourier transform, the constraint can be written as
\begin{equation*}
	- \frac{2\pi G}{3 V } \frac{p_a^2}{a} + \frac{1}{a} \int \tilde H\left[\vec k, \vec E(\vec k), \vec A(\vec k)\right] d^3\vec k = 0.
\end{equation*}
If we multiply it by $a$ the constraint splits
\begin{equation*}
	- \frac{2\pi G}{3 V } p_a^2 + \int \tilde H\left[\vec k, \vec E(\vec k), \vec A(\vec k)\right] d^3\vec k = 0.
\end{equation*}
The radiation part, in turn, splits into a sum over modes. Assuming a weak perturbation that allows interaction between the modes, the system matches the framework of section \ref{section_thermodynamics}, with $I_{\vec k} = \tilde H\left[\vec k, \vec E(\vec k), \vec A(\vec k)\right]$ and $I\equiv\sum_{\vec k} I_{\vec k} = -I_a$ fixed. Then, neglecting quantum aspects, one gets equipartition for $I$, i.e., Rayleigh-Jeans distribution
\begin{equation*}
	dI_\nu = 8\pi k_B T_I \nu^2 d\nu.
\end{equation*}
where $T_I$ is a conserved generalized temperature. The split of the constraint generates evolution in thermal time.  This is scaled with respect to proper time because proper-time evolution is generated by the conventional-energy $H=\frac{I}{a}$. The relation between the time-independent  temperature $T_I$ and the conventional temperature is $T= \frac{T_I}{a}$, which gives the time dependence of the temperature of the radiation dominated FLRW universe.


\end{document}